\documentclass[preprint,aps]{revtex4-1}
\usepackage[dvips]{graphicx}  % for graphics
\usepackage{amsmath,amsfonts}
\usepackage{bm}
\newcommand{\romanNum}[1]{\@roman{#1}}
\usepackage{color}  % for color
\def\be{\begin{eqnarray}}
\def\ee{\end{eqnarray}}
\def\beq{\begin{equation}}
\def\eeq{\end{equation}}

\def\({\left (}
\def\){\right )}
\def\[{\left [}
\def\[{\right ]}

\bmdefine{\bmk}{\bm{k}} \bmdefine{\bmx}{\bm{x}}
\bmdefine{\bmA}{\bm{A}} \bmdefine{\bmB}{\bm{B}}
\bmdefine{\bmJ}{\bm{J}}

\newcommand{\calO}{\mathcal{O}}
\begin{document}

\author{Hua-Bi Zeng$^{1}$  Wei-Min Sun$^{1,2}$ and Hong-Shi Zong$^{1,2}$}
\address{$^{1}$ Department of Physics, Nanjing University, Nanjing 210093, China}
\address{$^{2}$ Joint Center for Particle, Nuclear Physics and Cosmology, Nanjing 210093, China}
\title{Supercurrent in $p$-wave Holographic Superconductor }

\begin{abstract}
The $p$-wave and $p+ip$-wave holographic superconductors with fixed DC supercurrent are studied by
introducing a non-vanishing vector potential. We find that close to the critical temperature $T_c$ of
zero current, the numerical results of both the $p$ wave model and the $p+ip$ model are the
same as those of Ginzburg-Landau (G-L) theory, for example, the critical current $j_c \sim (T_c-T)^{3/2}$ and
the phase transition in the presence of a DC current is a first
order transition. Besides the similar results between
both models, the $p+ip$ superconductor shows isotropic behavior for the supercurrent,
while the $p$-wave superconductor shows anisotropic behavior for the
supercurrent.

\end{abstract}
\maketitle

\section{introduction}
The correspondence between gravity theory and quantum field theory
(gravity/gauge duality) \cite{1,2,3,4} has provided a novel approach to study the strongly coupled systems in condensed matter physics. This duality is a weak/strong duality, which means that we can study a strongly coupled field theory on the boundary by studying a weakly coupled quantum gravity theory on the bulk. Furthermore, in the large $N$ limit we need only classical gravity in the bulk. One important application of AdS/CM
is the holographic superconductors. The holographic superconductors can be
realized because the fact that an asymptotic AdS black hole coupled with matter fields will have hair
below a critical temperature and consequently it will be superconducting \cite{5}.
While above the critical temperature, the black hole has no
hair, which corresponds to the normal state.

There are several holographic models of superconductor with different matter sectors.
The model of $s$-wave holographic superconductor contains a black hole, a charged scalar field coupled to a Maxwell
field\cite{6,7,8,9,10}.
In Ref. \cite{36}, the scalar field condensation instability of rotating anti-de Sitter black
holes was also studied. If the charged scalar field
in the $s$-wave model is replaced by a charged symmetric traceless tensor field, we will get a $d$-wave holographic superconductor \cite{11,12,13,14,15}.
The phases of an $s$-wave holographic superconductor with
fixed superfluidity velocity were studied in Refs. \cite{29,30,31}, while the phases with fixed DC current was studied in Ref. \cite{30,32}. It is found that the
$s$-wave holographic superconductor with DC current shows the same results as the
G-L theory for superconducting films, for example, the phase transition at $T_c$ in the presence of a
DC current becomes a first order transition, the critical current $j_c \sim (T_c-T)^{3/2}$ close to $T_c$.

The $p$-wave holographic superconductor composed of non-Abelian gauge fields (the matter
sector) and a black hole background (the gravity sector) was first studied in Ref. \cite{16}.
The appearance of superconductivity is due to the
condensate of non-Abelian gauge fields in the theory.
The properties of this non-Abelian holographic superconductor has been
studied in Refs. \cite{17,18,19,20,21,22,23,24,25,26,34,27,28} (for a review one can see Ref. \cite{35}).
The model with fixed superfluidity velocity has also been studied in Ref. \cite{22}. It is found that the phase transition at small velocity is
a second order transition while at large velocity it is a first order
transition. This is similar to the $s$-wave model with fixed velocity studied in Refs. \cite{29,30,31}.
With the familiarity between these two holographic superconductor models with fixed velocity, we expect that the results similar to those of an $s$-wave model with DC currents can also appear in the $p$-wave model. Since the study of $p$-wave holographic superconductor with fixed DC current is lacking,
we study this issue in this paper. The results we \emph{obtain} are: (a) At any fixed DC current the superconducting
phase transition is a first order transition. (b) At a fixed temperature close to $T_c$, the $j_y$ versus $v_y$ curve has exactly the same features as that of the G-L
theory. (c) The critical current $j_c \sim (T_c-T)^{3/2}$ near $T_c$.
(d) Near $T_c$, the squared ratio of the maximal condensate to the minimal condensate is equal to two thirds at a fixed temperature;
the maximal value of condensation corresponds to the zero current while the minimal value of condensation corresponds to the critical current. These results are also similar to those of the $s$-wave model \cite{30,32}.
In this paper we also study the $p+ip$ wave holographic superconductor with DC current, which
shows results similar to the $p$-wave one.
However, the supercurrent in the $p+ip$ wave model is isotropic, and
the $p$-wave superconductor shows anisotropic behavior for the
supercurrent. The anisotropic behavior of the supercurrent
in the $p$-wave superconductor also differentiates the $p$-wave superconductor from the $s$-wave superconductor.

The organization of this paper is as follows.
In section II, we first review the dual gravity theory of the $p$-wave superconductor
with DC current along the $y$ direction and the equations of motion are also given. Then by numerically solving these equations, we obtain our results, such as the relation between the condensation and the temperature and the phase diagram. A comparison of these results to that of G-L theory and the $s$-wave model is also presented.
In section III, we give some results for the $p+ip$-wave superconductors with
DC current along the $y$ direction, which are similar to that of the $p$-wave model.
The study of the $p$ wave and $p+ip$ wave holographic superconductors with a DC current $j_x$ in the $x$ direction is
given in section IV.
Finally, the discussion and conclusions are presented in section V.

\section{$p$-wave holographic superconductor with current}
In this section, we first review the action of the
non-Abelian holographic superconductor and give the equations of motion (EOMs) of the
model with a nonvanishing vector potential. Then, after numerically solving the EOMs, we give the results and their physical meanings.

\subsection{The Dual Gravity Theory}
The action of the $p$-wave holographic superconductor includes the
Einstein-Hilbert action and an SU(2) gauge field, which is called the Einstein-Yang-Mills (EYM) theory. It has the following action \cite{5}
\begin{eqnarray}
S_{\textmd{EYM}}=
\int\sqrt{-g}d^4x\left[\frac{1}{2\kappa_4^2}\left(R+\frac{6}{L^2}\right)
-\frac{L^2}{2g_{\rm YM}^2}\textmd{Tr}(F_{\mu\nu}F^{\mu\nu})\right],
\end{eqnarray}
where $g_{\rm YM}$ is the gauge coupling constant and
$F_{\mu\nu}=T^aF^a_{\mu\nu}=\partial_\mu A_\nu-\partial_\nu
A_\mu-i[A_\mu,A_\nu]$ is the field strength of the gauge field
$A=A_\mu dx^\mu=T^aA^a_\mu dx^\mu$. For the $SU(2)$ gauge symmetry,
$[T^a,T^b]=i\epsilon^{abc}T^c$ and
$\textmd{Tr}(T^aT^b)=\delta^{ab}/2$, where $\epsilon^{abc}$ is the
totally antisymmetric tensor with $\epsilon^{123}=1$. The Yang-Mills
Lagrangian becomes
$\textmd{Tr}(F_{\mu\nu}F^{\mu\nu})=F^a_{\mu\nu}F^{a\mu\nu}/2$ with
the field strength components $F^a_{\mu\nu}=\partial_\mu
A^a_\nu-\partial_\nu A^a_\mu+\epsilon^{abc}A^b_\mu A^c_\nu$.

Working in the probe limit in which the matter fields do not
backreact on the metric as in Refs. \cite{16,17,18}
and taking the planar Schwarzchild-AdS ansatz, the  black hole
metric reads (we use mostly plus signature for the metric)
\begin{equation}
ds^2=-f(r)dt^2+\frac{dr^2}{f(r)}+\frac{r^2}{L^2}(dx^2+dy^2),
\label{metric}
\end{equation}
where the metric function $f(r)$ is
\begin{equation}
f(r)=\frac{r^2}{L^2}(1-\frac{r_0^3}{r^3}).
\end{equation}
$L$ and $r_0$ are the radius of the AdS spacetime and the horizon
radius of the black hole, respectively. We can set $L=1$. Then the Hawking
temperature of the black hole reads
\begin{equation}
T=\frac{3r_0}{4\pi},
\end{equation}
which is also the temperature of the dual gauge theory living on the
boundary of the AdS spacetime.

It is convenient to introduce a new coordinate $z=1/r$. The metric (\ref{metric}) then becomes
 \begin{equation}
ds^2=\frac{1}{z^2}(-h(z)dt^2+dx^2+dy^2+\frac{dz^2}{h(z)}),
 \end{equation}
where $h(z)=1-(z/z_h)^3$ and $z_h=1/r_0$ is the horizon. For convenience we set $z_h=1$ in our calculation.

Using the Euler-Lagrange equations, one can obtain the equations of
motion for the gauge fields,
\begin{equation}
\frac{1}{\sqrt{-g}}\partial_{\mu}\left(\sqrt{-g}F^{a\mu\nu}\right)
+\epsilon^{abc}A^{b}_{\mu}F^{c\mu\nu}=0.
\end{equation}
For the $p$-wave backgrounds, in order to study
the DC current of the model, we need a non-vanishing vector potential.
Then the ansatz takes the
following form,
\begin{equation}
A=\phi(z)T^3dt+A_y^{3}(z)T^3dy+w(z)T^1dx.
\label{ansatzp}
\end{equation}
Here the $U(1)$ subgroup of $SU(2)$ generated by $T^3$ is identified
to the electromagnetic gauge group \cite{16} and $\phi$ is
the electrostatic potential, which must vanish at the horizon for
the gauge field in order for $\phi dt$ to be well-defined, but need not vanish at
infinity. Thus the black hole can carry charge through the
condensation of $w$, which spontaneously breaks the $U(1)$ gauge
symmetry. This is a Higgs mechanism.
With this ansatz (\ref{ansatzp}), we can derive the equations of motion,
\begin{equation}
(-1+z^3)^2 \frac{d^2w(z)}{dz^2}+3z^2(-1+z^3) \frac{d w(z)}{dz}+\phi^2(z)w(z)+(-1+z^3)(A_y^3(z))^2w(z)=0,
\label{eom1}
\end{equation}

\begin{equation}
(-1+z^3)\frac{d^2\phi''(z)}{dz^2}+\phi(z)w^2(z)=0,
\label{eom2}
\end{equation}
and
\begin{equation}
(-1+z^3)\frac{d^2 A_y^3(z)}{dz^2}+3z^2\frac{d A_y^{3}(z)}{dz}+A_y^3(z)w^2(z)=0.
\end{equation}

In order to solve these equations we need to specify the boundary
conditions on both the boundary and the horizon. On the horizon $z=z_h$,
the scalar potential $\phi$ should vanish at the horizon in order to
make $\phi dt$ well defined. At the horizon the fields $w$ and $A_y^3$ should be regular. On the boundary, the asymptotic behaviors of the three fields take the
following form

\begin{equation}
  w=\frac{\langle
\cal O\rangle}{\sqrt{2}}z
  + \cdots,
\end{equation}

\begin{equation}
 \phi = \mu - \rho z + \cdots,
\end{equation}

\begin{equation}
 A_y^3=v_y-j_y z+\cdots.
\end{equation}
From the AdS/CFT dictionary we explain $\mu$ as the chemical potential, $\rho$
as the charge density, $v_y$ the superfluid velocity and $j_y$ the supercurrent
along the $y$ direction of the boundary field theory. $\langle
\calO\rangle$ is the order parameter of the superconducting phase. The constant term in
Eq. (II.11) is set to zero by requiring that there be no source term for the operator $\langle \calO \rangle$ in the field theory action.

Before solving these equations, let us consider the non-superconducting state with $w=0$. Then the solution of $\phi$ is $\phi=\mu(1-z)$. The equation for $A_y^3$ becomes
\begin{equation}
(-1+z^3)\frac{d^2 A_y^3(z)}{dz^2}+3z^2\frac{d A_y^{3}(z)}{dz}=0.
\end{equation}
The solution of this equation takes the following form
\begin{equation}
A_y^3(z)=c_1+c_2(-\frac{\arctan(\frac{1+2z}{\sqrt3})}{\sqrt 3}+\frac{1}{3}\log(z-1)-\frac{1}{6}\log(1+z+z^2)),
\end{equation}
where $c_1$ and $c_2$ are two constants.
$c_2$ must equal to zero since we require that the energy density near the horizon be finite.
If $c_2$ is non-vanishing, then near the horizon the term with $\log(z-1)$ gives a contribution to $F_{z,y}^3 \sim \partial_zA_y^3\sim c_2 \frac{1}{z-1}$.
Then the energy density near the horizon has a contribution from $g^{zz}g^{yy}F_{zy}^2$ which diverges as $1/(z-1)$ at the horizon $z_h=1$. This unphysical behaviors means that there is no supercurrent for the non-superconducting state since $A_y^3$ must be a
constant. However, the non-superconducting state can have a superfluidity with value $c_1$. We will come back to this issue again when we
calculate the free energy later.

\subsection{Order Parameter via Temperature}
To study the behaviors of a $p$-wave model with supercurrent means that
we have to solve the equations of motion
with fixed $j_y$. We can also solve the
equations with a fixed $v_y$, which corresponds to
studying the phases of a superconductor with a fixed superfluidity velocity.
This has been done in Ref. \cite{22}.
The first important problem to study is how the order parameter changes with the temperature for
this holographic superconductor with current.

\begin{figure}
\includegraphics[width=6cm,clip]{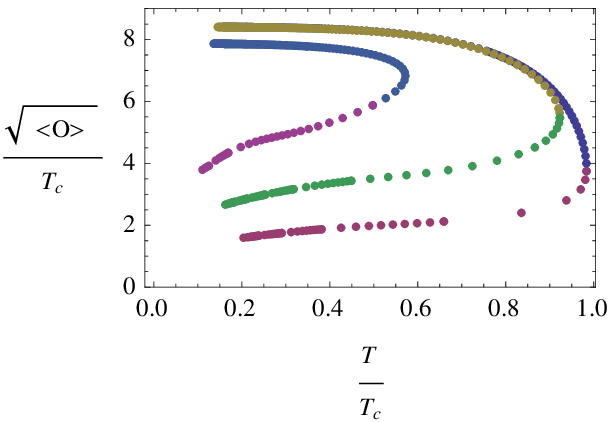}
\includegraphics[width=6cm,clip]{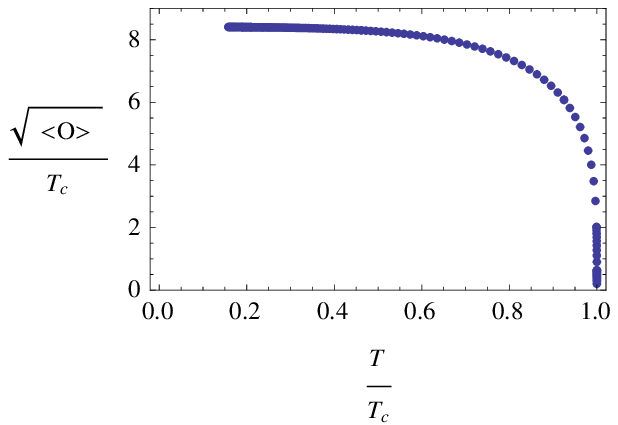}
\caption{Plot of the order parameter versus the temperature for different values of the
current. The left panel shows three different plots of $\langle
\calO\rangle$ versus $T$ for $j_y$=1/10, 1/100, 1/1000 (from the right to the left). The right panel shows the plot of $\langle\calO\rangle$ versus $T$ with zero current, it is clear that the phase transition is a second order transition now.}
\end{figure}

From Fig. 1 it can be seen that when the supercurrent is not zero, there are two solutions of the order parameter corresponding to a fixed temperature. In the next
section we will show that the solution with lower value of the order parameter takes
a larger free energy than the solution with larger values of the order parameter and therefore
it is unfavorable. The critical temperature decreases as
the current increases, which indicates that there should exist a critical current
above which there is no superconductivity. When one
lowers the temperature from a temperature above the critical one, the order of phase transition
at the critical temperature for a fixed current should be a first order one, since the order parameter jumps from zero to a finite value at the critical temperature. Such a jump will certainly change the energy and so requires some latent heat, which
implies that the phase transition should be a first order one. This conclusion is the same as the one we shall give from observing the curve of the current $j_y$ versus the superfluid velocity $v_y$ at a fixed temperature. This result is different from the one obtained for the case of a fixed superfluidity velocity, where at small velocity the phase transition is still a second order one. When the velocity becomes larger, the phase transition turns to be of first
order \cite{22}. For the $s$-wave holographic superconductor with current,
the order parameter is also bivaluated, and the states with lower value of the condensate have a larger free energy than
their counterparts with larger values of the condensate at the same temperature. This is the same as the result of a $p$-wave
holographic superconductor.

\subsection{The Free Energy}
To confirm that compared with the state with
larger parameter, the state with lower parameter is unfavorable, we need to compare their free energies.
The free energy of the field theory is determined by the value of
the Yang-Mills action (ignoring the back-reaction of the gauge fields on the metric)
\begin{equation}
S_{\textmd{YM}}= \int d^4x\mathcal{L}_{YM}
\end{equation}
evaluated on-shell up to boundary counterterms,
$F=-TS_{\textmd{os}}+\cdots$, where the ellipsis denotes boundary
terms that we should introduce to regulate the action when needed.
The on-shell Yang-Mills action $S_{\textmd{os}}$ is determined by
plugging the equations of motion, Eq. (II.8), Eq. (II.9) and Eq. (II.10) into the explicit
form of the Yang-Mills Lagrangian (omitting the irrelevant
factor $1/4g_{YM}^2$)
\begin{equation}
\begin{array}{ll}
S_{\textmd{os}}=\int{d^3x}(-\phi\phi'+2z^2fww'-A_y^3 (A_y^{3})' ) |_{z=\epsilon}
-\int{d^3x}\int^{z_h}_{\epsilon}dz
\left(\frac{\phi^2w^2}{1-z^3}+w^2(A_y^3(z))^2\right),
\end{array}
\end{equation}
where we have used the coordinate $z=1/r$, and $z =\epsilon = 0^+$
is the boundary of the AdS spacetime.

To regulate $S_{\textmd{os}}$, it is important to choose an
ensemble. By keeping $\mu$ fixed, we are working in the grand
canonical ensemble without an additional boundary term. Near the boundary $z=\epsilon$, the fields $\phi$
and $w$ are determined by Eq. (II.11), Eq. (II.12) and Eq. (II.13) and the three terms $-\phi
\phi'$, $-A_y^3 (A_y^{3})'$ and  $2z^2fww'$ in Eq. (II. 17) give $\mu \rho$, $v_y j_y$ and $2w_0w_1$,
respectively. We can see that the on-shell action
$S_{\textmd{os}}$ is not divergent and no counterterms are
needed. Since $w_0$ is fixed to be zero, for a spatially
homogenous system, the free energy density of the field theory
takes the following form
\begin{equation}
F/V= -\mu\rho-v_y j_y
+\int^{z_h}_{\epsilon}dz\left(\frac{\phi^2w^2}{1-z^3}+w^2(A_y^3(z))^2\right),
\end{equation}
where $V\equiv\int{d^3x}$.

In Fig. 2 we present the free energy of a fixed current $j_s=1/100$ for the two branches of
solution. It can be clearly seen that the solution with a larger value of the order parameter has a lower free energy.

\begin{figure}
\includegraphics[width=6cm,clip]{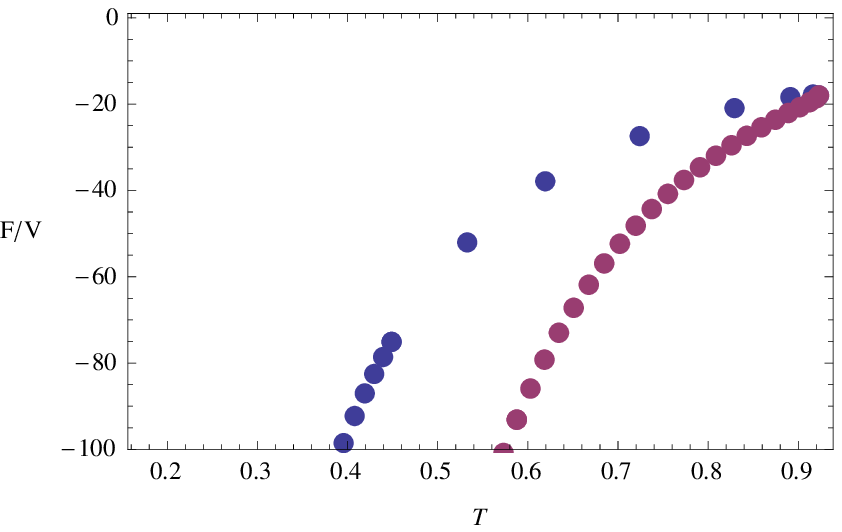}
\includegraphics[width=6cm,clip]{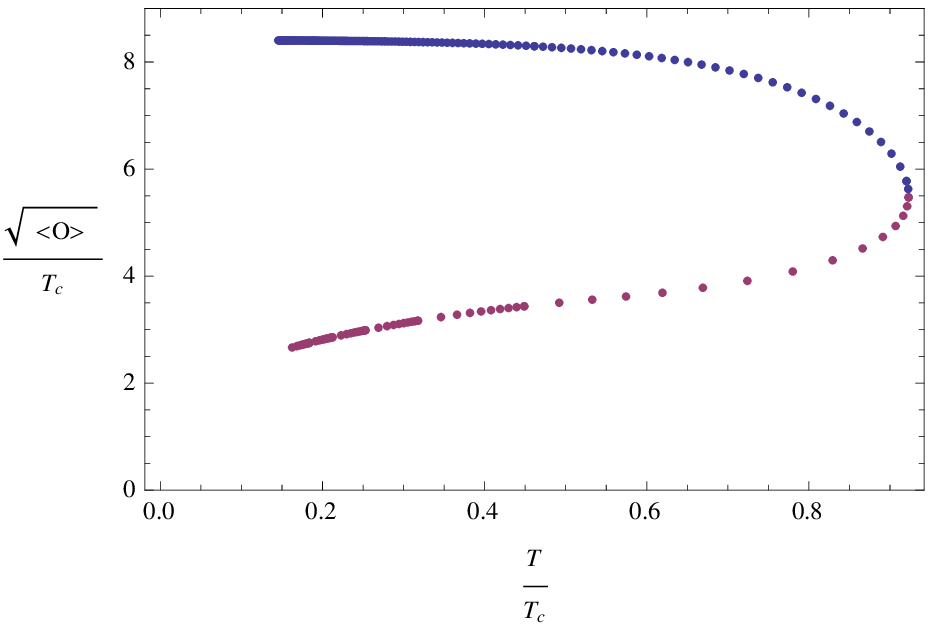}
\caption{Plot of the free energy of the superconducting
phases for $j_y=1/100$. The blue dotted line corresponds to the points with a lower value of the
condensate at a given temperature. The right panel shows the corresponding plot of the condensation versus $T$. It can be seen that the lower branch corresponds indeed to states
with larger free energy and is thus metastable.}
\end{figure}

Of course, the best way to show that the phase transition at the critical
temperature is of first order is to compare the free energy between the superconducting
state and the non-superconducting state with current. However, this is not possible in this model, since there is no non-superconducting state with a fixed current as is discussed at the end of Section A. Nevertheless,
with the fact that the order parameter goes discontinuously at the phase transition
point, there should be no problem to conclude that the phase transition is a first order one.

\subsection{Current via the Superfluidity Velocity}

Another physical quantity by which one can compare the difference between the gravity model of superconductor and the G-L theory is the relation between the current and the superfluidity velocity at a fixed temperature. From this relation we can also get the information of the
phase transitions at the critical current or critical velocity.

\begin{figure}
\includegraphics[width=6cm,clip]{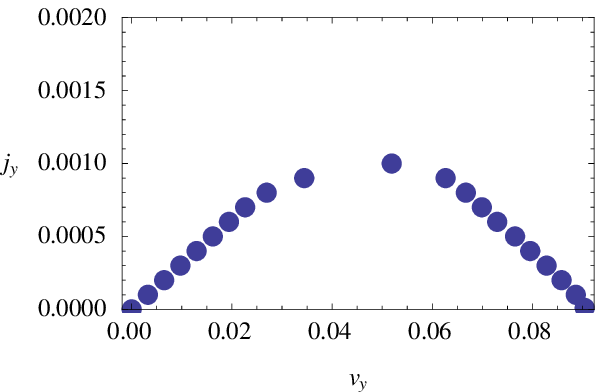}
\includegraphics[width=6cm,clip]{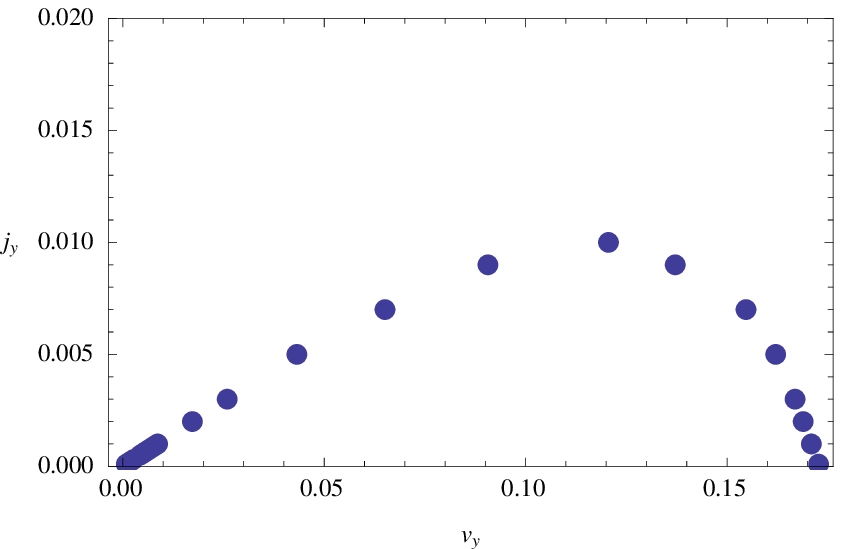}
\includegraphics[width=6cm,clip]{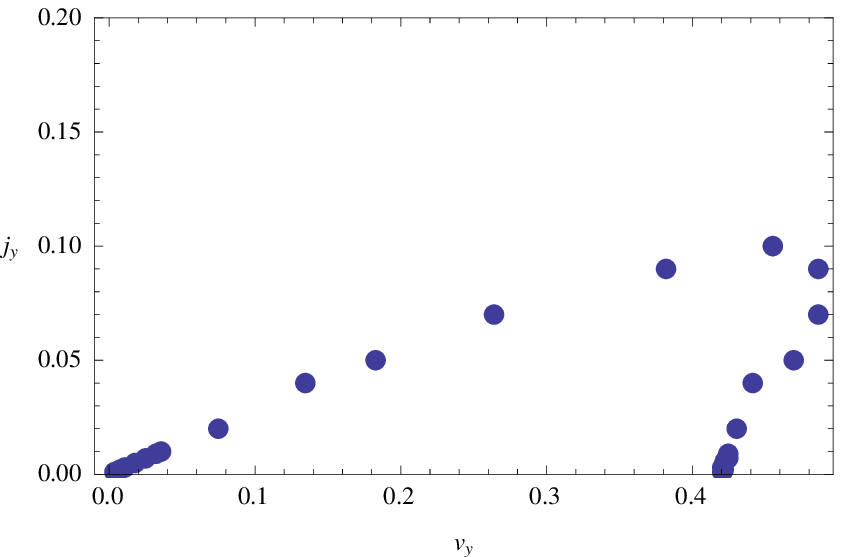}
\caption{Plots of the current $j_y$ versus the superfluid velocity $v_y$ at a fixed temperature.
The two panels above correspond to $T/T_c=0.9836, 0.9229$ (from left to right),
at which the critical currents $j_c$ are $1/1000$ and $1/100$, respectively.
The panel below shows the result for $T/T_c=0.4836$ corresponding to $j_c=1/10$.}
\end{figure}

The first two plots in Fig. 3 correspond to the temperature $T/T_c=0.9836, 0.9229$. For these two temperatures the critical current are $j_c=1/1000$ and $j_c=1/100$,
respectively. It can be clearly seen that for temperatures close to $T_c$,
where the G-L theory works very well, the plots of $j_y$ versus $v_y$ is the same
as that of G-L theory. From these plots we can also know the order of the phase transitions
at critical current or critical velocity. For the plots of $T/T_c=0.9836, 0.9229$, the maximal velocity corresponds to a vanishing
current, which means that the phase transitions at critical velocities $j_c=1/1000$ and $j_c=1/100$ are of
second order.

The third plot corresponding to a larger temperature $T/T_c=0.4836$ is different. The maximal velocity corresponds to a non-vanishing
current, which means that the phase transition at the critical velocity is a
first order one. So we can conclude that for small critical velocity the phase transition is a second order transition, while for large enough velocity the phase transition becomes a first order one. These are consistent with the model with a $p+ip$ background at a fixed superfluidity
velocity \cite{22}.  We can also check our results by
the curve of the order parameter versus the temperature for a fixed velocity (rather than a fixed current), which is plotted in Fig 4. From Fig. 4 it can be seen that for a small value of the current $v_y=0.1$, the condensation goes continuously at the critical temperature, which means that the phase transition is a second order one. While for a larger value of the current $v_y=0.48$,
which corresponds to the maximal value of the current in the third curve of Fig. 3, the condensation goes discontinuously at the critical temperature. This indicates that the phase transition is a first order one. According to the computation of the free energy in Refs. \cite{29,22}, it shows that the curve of the condensation versus the temperature, such as the right one, is indeed the sign of a first order phase transition.
An interesting thing is that all the results we obtain above are similar to that of the $s$-wave model \cite{30,32}.

\begin{figure}
\includegraphics[width=6cm,clip]{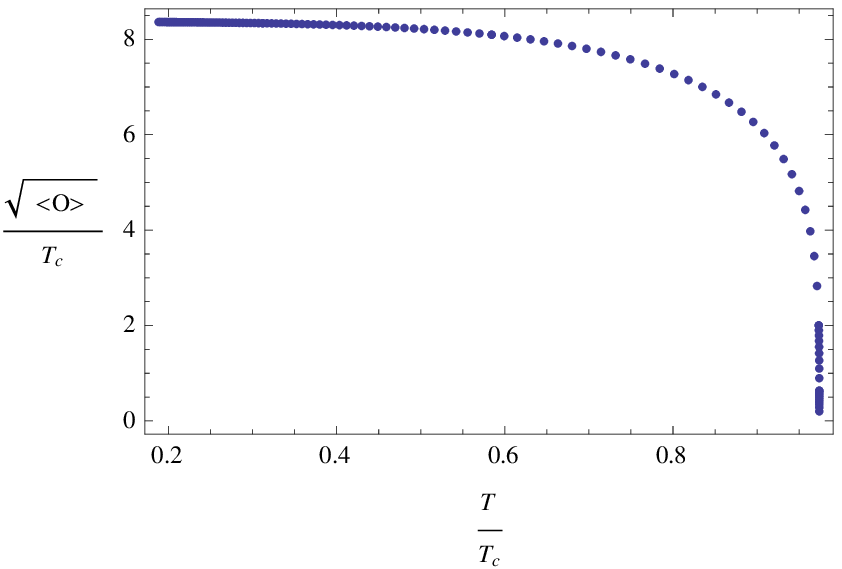}
\includegraphics[width=6cm,clip]{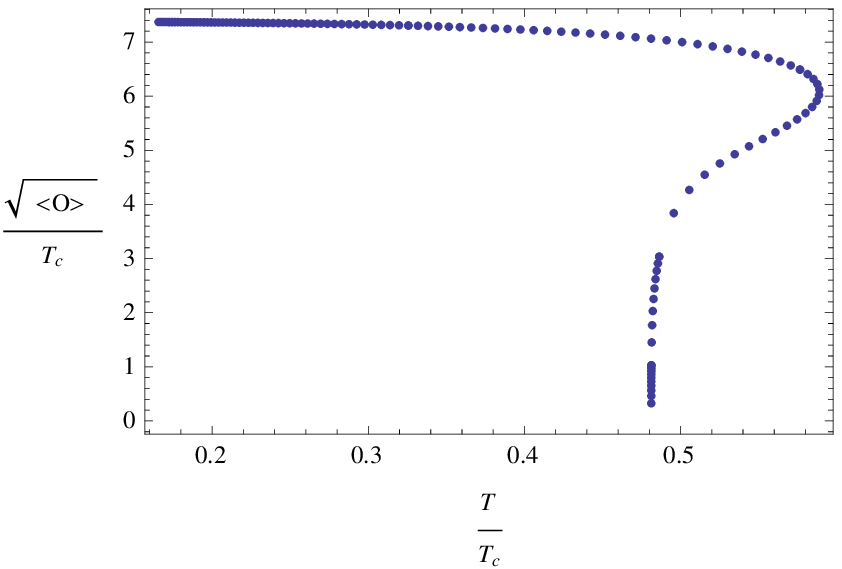}
\caption{Plot of the condensation versus the temperature for a fixed velocity. The left panel corresponds to $v_y=0.1$ and the right panel corresponds to $v_y=0.48$.}
\end{figure}

\subsection{The Critical Current via Temperature}

In this subsection we study the critical current $j_c$
for different $T$ near $T_c$ to compare the results with that of
the G-L theory. As predicted by G-L theory, $j_c$ is proportional to $(T_c-
T)^{3/2}$ when the temperature is close to $T_c$. As illustrated in Fig. 5, this scaling behavior is indeed obeyed by holographic
superconductors for temperatures close to $T_c$, and this is also the case in the $s$-wave model \cite{32}. Another prediction of the G-L theory is that, at any fixed temperature, the
norm of the condensate decreases monotonically with the velocity from
its maximal value, the maximal value $\langle \calO \rangle_\infty$ corresponding to zero velocity and zero current. As is shown in Fig. 3, the critical current is reached before the velocity reaches its maximal value. The norm of the condensate has an intermediate value $\langle \calO \rangle_c$ at the maximal current.
The G-L theory tells us that the squared ratio of $\langle \calO \rangle_c$ to the maximal condensation  $\langle \calO \rangle_\infty$
is exactly equal to $2/3$. From Fig. 6 it can be seen that this is indeed the same case for the $p$-wave holographic superconductor.

\begin{figure}
\includegraphics[width=6cm,clip]{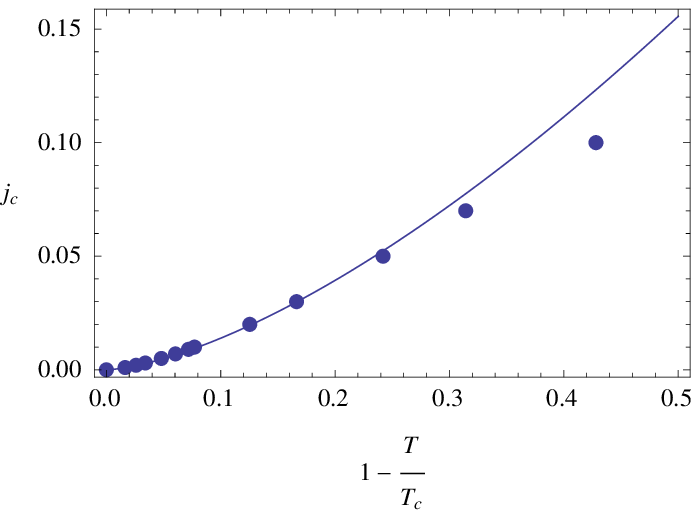}
\includegraphics[width=6cm,clip]{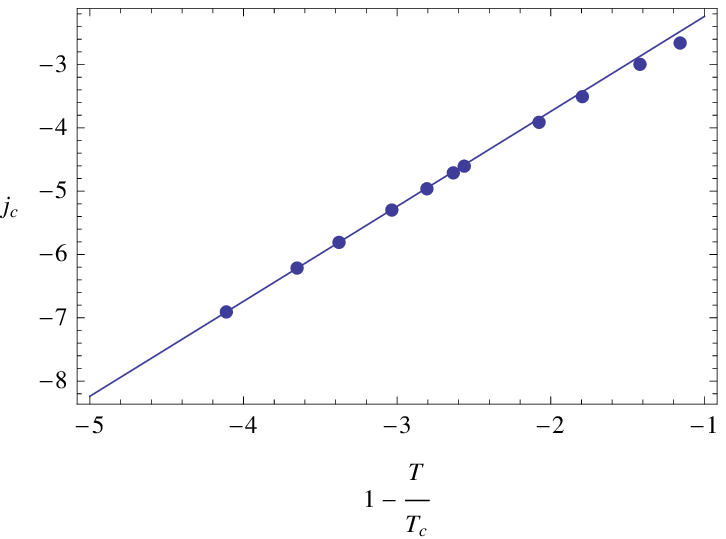}
\caption{ Plot of the critical current versus the temperature. The right panel shows a
log-log plot from which we can read off the critical exponent, getting 1.497, which agrees
with the expected GL scaling of 3/2 within numerical precision. The left panel shows
the $j_c$ versus $(1-T/T_c)$, the solid line is $0.44(1-T/T_c)^{3/2}$.}
\end{figure}

\begin{figure}
\includegraphics[width=6cm,clip]{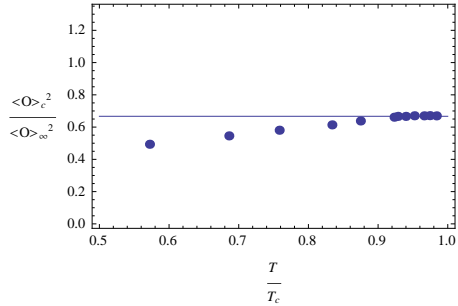}
\caption{ Plot of the ratio ($\langle \calO \rangle_c / \langle \calO \rangle_\infty)^2$ versus the temperature. The solid line
corresponds to the value of 2/3 predicted by the G-L theory and it also appears in the $s$-wave model.}
\end{figure}

\section{the $p+ip$ backgrouds with current}
The EYM theory with a $p+ip$ background has the following ansatz \cite{17,18},
\begin{equation}
A=\phi(z)T^3dt+A_y^{3}(z)T^3dy+w(z)T^1dx+w(z)T^2dy.
\end{equation}

With this ansatz, the EOMs are
\begin{eqnarray}
&&(-1+z^3)^2 \frac{d^2w(z)}{dz^2}+3z^2(-1+z^3) \frac{d w(z)}{dz}+(-1+z^3) w^3(z)\nonumber\\
&&+\phi^2(z)w(z)+\frac{(-1+z^3)(A_y^3(z))^2 w(z)}{2}=0,
\label{eom1}
\end{eqnarray}
\begin{equation}
(-1+z^3)\frac{d^2\phi''(z)}{dz^2}+2\phi(z)w^2(z)=0,
\label{eom2}
\end{equation}
and
\begin{equation}
(-1+z^3)\frac{d^2 A_y^3(z)}{dz^2}+3z^2\frac{d A_y^{3}(z)}{dz}+A_y^3(z)w^2(z)=0.
\end{equation}
The boundary conditions for the fields $w$, $\phi$ and $A_y^3$ are the same as those in the case of $p$-wave background discussed in Section II.A.

After solving the EOMs numerically, we find similar results as those of the
$p$-wave holographic superconductor discussed in last section. These results are shown in Figs. 7-9.

Until now, for both $p$ wave and $p+ip$ wave backgrounds of non-Abelian holographic
superconductor with DC current along the $y$ direction, we get results extremely close to those of
the $s$-wave model. A first look of the EOMs of both $p$ wave and $p+ip$ wave holographic superconductors
with DC current makes us find that the Eq. (II. 8), Eq. (II. 9), Eq. (II. 10), Eq. (III. 20), Eq. (III. 21) and Eq. (III. 22) are
different from the corresponding ones of the $s$-wave model in Ref. \cite{30}. For the non-Abelian holographic superconductors the expansion of the field $w(z)$ goes as a
constant plus a term linear in $z$ (equation II.11). In Ref. [18], their
scalar field $\psi(z)$'s expansion goes as a linear term plus a term
quadratic in $z$ (equation (10)). If we do a field redefinition, such
that $\psi(z) = z w(z) / \sqrt{2} $ from the scalar field of Ref. [18],
Eq. (II.8) for $w(z)$ and the corresponding one in Ref. [18] will differ by a
single term proportional to
\begin{equation}
z (-1 + z^3) w(z).
\end{equation}
But the equations for the scalar potential and the vector potential are the same as the corresponding ones in Ref. [18].
According to our computation, this term (Eq. (III.23)) will not affect the qualitative results but will change the quantitative results. For the $p+ip$ wave model, the situation is
different due to the $w^3$ term in Eq. (III.20). In the next section we will turn
on $j_x$ rather than $j_y$ for both $p$ wave and $p+ip$ wave backgrounds, and we will see that these two backgrounds are different with a nonvanishing $A_x$.

\begin{figure}
\includegraphics[width=6cm,clip]{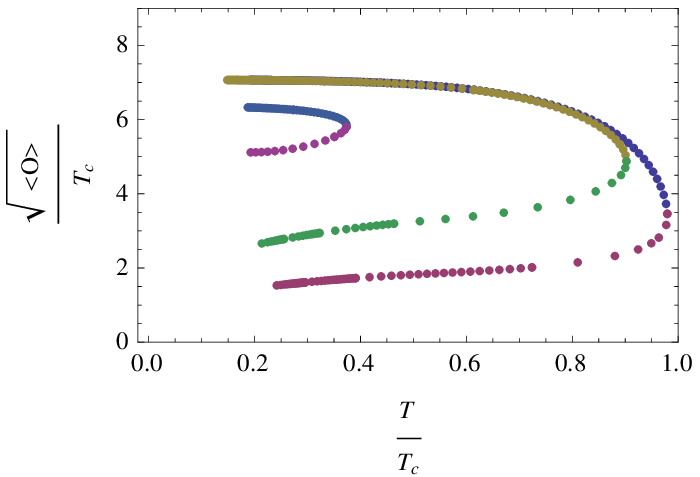}
\includegraphics[width=6cm,clip]{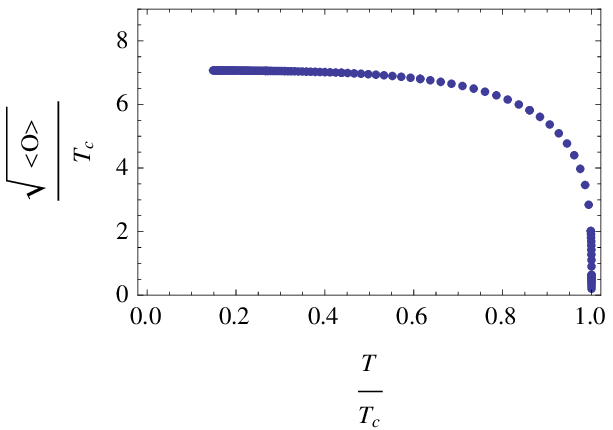}
\caption{Plot of the order parameter versus the temperature for different values of
current for the case of $p+ip$ background. The left panel shows the three different plots of
$\langle\calO\rangle$ versus the temperature for $j_y$=1/10, 1/100, 1/1000 (from the right to the left). The right panel shows the plot of $\langle
\calO\rangle$ versus the temperature with zero current. }
\end{figure}

\begin{figure}
\includegraphics[width=6cm,clip]{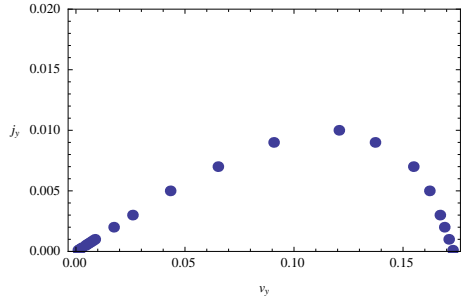}
\caption{Plot of the current $j_y$ versus the superfluid velocity $v_y$ for $p+ip$ background at a fixed temperature $T/T_c=0.9284$, at which
the critical current is $1/100$. }
\end{figure}

\begin{figure}
\includegraphics[width=6cm,clip]{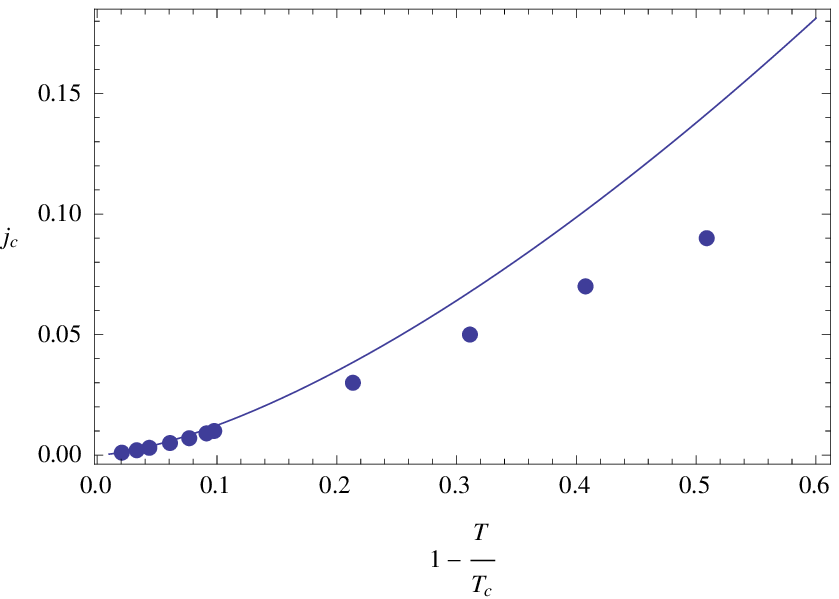}
\includegraphics[width=6cm,clip]{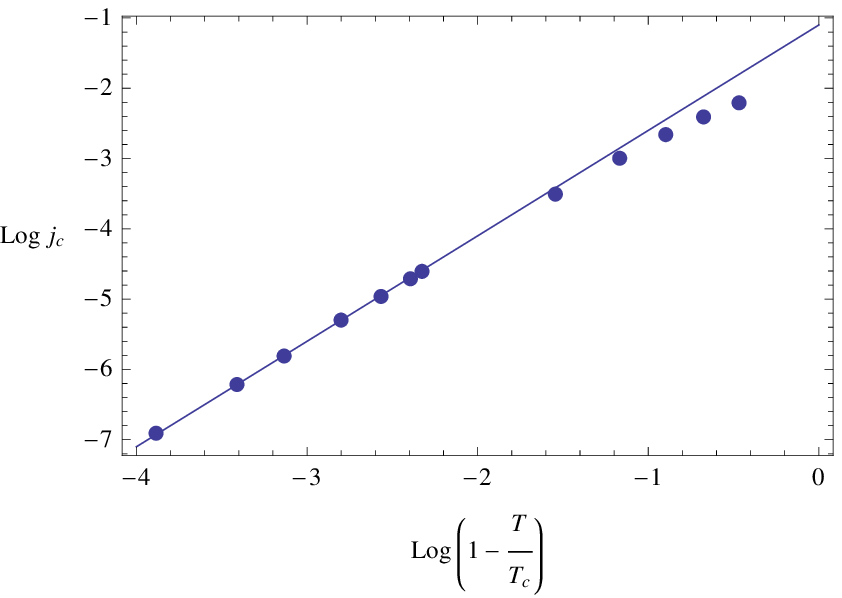}
\caption{ Plot of the critical current versus the temperature for $p+ip$ background. The right panel shows a
log-log plot from which we can read off the critical exponent, getting 1.498, which agrees
with the expected G-L scaling of 3/2 within numerical precision. The left panel shows
the $j_c$ versus $(1-T/T_c)$, the solid line is $0.39(1-T/T_c)^{3/2}$.}
\end{figure}

\section{Currents along the $x$ direction}
Now let us study whether there is any difference between the $p$-wave, $p+ip$ and $s$-wave model with DC current.
For the $p+ip$ wave background, by assuming a nonvanishing $A_x^3$, the EOMs become
\begin{eqnarray}
&&(-1+z^3)^2 \frac{d^2w(z)}{dz^2}+3z^2(-1+z^3) \frac{d w(z)}{dz}\nonumber\\
&&+(-1+z^3) w^3(z)+\phi^2(z)w(z)+\frac{(-1+z^3)(A_x^3(z))^2 w(z)}{2}=0,
\end{eqnarray}
\begin{equation}
(-1+z^3)\frac{d^2\phi''(z)}{dz^2}+2\phi(z)w^2(z)=0,
\end{equation}
and
\begin{equation}
(-1+z^3)\frac{d^2 A_x^3(z)}{dz^2}+3z^2\frac{d A_x^{3}(z)}{dz}+A_x^3(z)w^2(z)=0.
\end{equation}
We can see that these equations are the same as Eq. (III. 20), Eq. (III. 21) and Eq. (III. 22).
Therefore, the same results for $j_x$ will be expected.
We can also turn on a current along an arbitrary direction, which can
be done by turning on both $j_x$ and $j_y$ with a relationship
like $j_x=k j_y$. For example, a current along the direction of 45 degree to the $x$ direction corresponds to $k=1$. With both $j_x$ and $j_y$, the EOMs becomes
\begin{eqnarray}
&&(-1+z^3)^2 \frac{d^2w(z)}{dz^2}+3z^2(-1+z^3) \frac{d w(z)}{dz}+(-1+z^3) w^3(z)\nonumber\\
&&+\phi^2(z)w(z)+\frac{(-1+z^3)((A_x^3(z))^2+(A_y^3(z))^2) w(z)}{2}=0,
\end{eqnarray}
\begin{equation}
(-1+z^3)\frac{d^2\phi''(z)}{dz^2}+2\phi(z)w^2(z)=0,
\end{equation}
\begin{equation}
(-1+z^3)\frac{d^2 A_x^3(z)}{dz^2}+3z^2\frac{d A_x^{3}(z)}{dz}+A_x^3(z)w^2(z)=0.
\end{equation}
and
\begin{equation}
(-1+z^3)\frac{d^2 A_y^3(z)}{dz^2}+3z^2\frac{d A_y^{3}(z)}{dz}+A_y^3(z)w^2(z)=0.
\end{equation}

Since the equations for $A_x$ and $A_y$ are the same, $j_x=k j_y$ means that $A_x=k A_y$. Then the EOMs reduce to
\begin{eqnarray}
&&(-1+z^3)^2 \frac{d^2w(z)}{dz^2}+3z^2(-1+z^3) \frac{d w(z)}{dz}+(-1+z^3) w^3(z)\nonumber\\
&&+\phi^2(z)w(z)+\frac{(-1+z^3)(k^2+1)(A_y^3(z))^2 w(z)}{2}=0,
\end{eqnarray}
\begin{equation}
(-1+z^3)\frac{d^2\phi''(z)}{dz^2}+2\phi(z)w^2(z)=0,
\end{equation}
and
\begin{equation}
(-1+z^3)\frac{d^2 A_y^3(z)}{dz^2}+3z^2\frac{d A_y^{3}(z)}{dz}+A_y^3(z)w^2(z)=0.
\end{equation}
Now the value of current $j=j_y \sqrt{k^2+1} $. After a redefinition $A_y \longrightarrow A_y/\sqrt{k^2+1}$, the EOMs turn back to Eq. (III. 20), Eq. (III. 21) and Eq. (III. 22) with the same magnitude of DC current.
From this fact we conclude that if a current in both the $x$ and $y$ direction is
present, the equations of motion only depend on the magnitude of the
current and not on its direction. Then the supercurrent in
$p+ip$ holographic superconductor is isotropic, which is similar to the case of the $s$ wave model.

For the $p$ wave background with $j_x$, the EOMs are
\begin{equation}
(-1+z^3)^2 \frac{d^2w(z)}{dz^2}+3z^2(-1+z^3) \frac{d w(z)}{dz}+\phi^2(z)w(z)=0,
\label{eom1}
\end{equation}
\begin{equation}
(-1+z^3)\frac{d^2\phi''(z)}{dz^2}+\phi(z)w^2(z)=0,
\label{eom2}
\end{equation}
and
\begin{equation}
(-1+z^3)\frac{d^2 A_x^3(z)}{dz^2}+3z^2\frac{d A_x^{3}(z)}{dz}=0.
\end{equation}
From these equations it can be seen that $A_x$ is independent of
$w$. $A_x$ is totally determined by Eq. (IV.29). This equation is also the same as
Eq. (II.14). According to the study of Eq. (II.14) in
Section II. A, $A_x$ must be a constant. From this we conclude that there is no
supercurrent along the $x$ direction. Just as we have discussed for the $p+ip$ background, we
can also try to turn on a current along an arbitrary direction, which means that we have to turn on both $A_x^3$ and $A_y^3$. Then the EOMs turn to
\begin{equation}
(-1+z^3)^2 \frac{d^2w(z)}{dz^2}+3z^2(-1+z^3) \frac{d w(z)}{dz}+\phi^2(z)w(z)+(-1+z^3)(A_y^3(z))^2w(z)=0,
\label{eom1}
\end{equation}
\begin{equation}
(-1+z^3)\frac{d^2\phi''(z)}{dz^2}+\phi(z)w^2(z)=0,
\label{eom2}
\end{equation}
\begin{equation}
(-1+z^3)\frac{d^2 A_y^3(z)}{dz^2}+3z^2\frac{d A_y^{3}(z)}{dz}+A_y^3(z)w^2(z)=0.
\end{equation}
and
\begin{equation}
(-1+z^3)\frac{d^2 A_x^3(z)}{dz^2}+3z^2\frac{d A_x^{3}(z)}{dz}=0.
\end{equation}

It is also clear that $A_x$ is still independent of $w$, so there also should be no
$j_x$, as has been discussed. The current can only flow along the $y$ direction,
then the DC current in the $p$ wave holographic superconductor is
anisotropic.

\section{discussion and conclusion}
In this paper, we study the $p$-wave and the $p+ip$ wave holographic
superconductor with DC supercurrent. For the $p$ wave background with DC
current along the $y$ direction and the $p+ip$ wave background with DC current along both
$x$ and $y$ directions, the results near the
critical temperature agree quantitatively with several
properties of the Ginzburg-Landau theory. For example, the squared ratio
of the maximal condensate to the minimal condensate is equal to two thirds, the
critical current is proportional to $(T_c-T)^{3/2}$. While
for the $p$ wave model there is no supercurrent along the $x$ direction.
However, it is interesting to note that the non-Abelian holographic superconductors show
the same mean-field behaviors as the $s$-wave model, which are also the
results of the G-L theory. These results make us believe that the
holographic description of superconductors indeed contains some physics of real
world superconductors. Maybe an analytical analysis of these two models are helpful to explaining why the holographic models of $s$-wave superconductor and $p$-wave superconductor show similar results \cite{33,37}.

\section{acknowledgement}

We especially thank C. P. Herzog for helps on the numerics. We thank Xin Gao for many valuable comments. This work is supported in part by the National Natural Science
Foundation of China (under Grant Nos. 10775069, 10935001 and 11075075),
the China Postdoctoral Science Foundation (Grant No. 20100481120) and the Research Fund
for the Doctoral Program of Higher Education (Grant No. 20080284020).

\end{document}